\documentclass[%
reprint,
superscriptaddress,
 aps,
prstab,
floatfix
]{revtex4-2}

\usepackage{graphicx}
  \setkeys{Gin}{width=\linewidth}
\usepackage{dcolumn}
\usepackage{bm}

\usepackage{amsmath,amsfonts,amssymb}
\usepackage{amsbsy}

\usepackage{hyperref}
\hypersetup{
    colorlinks=true,       
    linkcolor=blue,          
    citecolor=blue,        
    filecolor=blue,         
    urlcolor=blue        
}

\newcommand{\orcidicon}[1]{\href{https://orcid.org/#1}{\includegraphics[height=\fontcharht\font`\B,keepaspectratio]{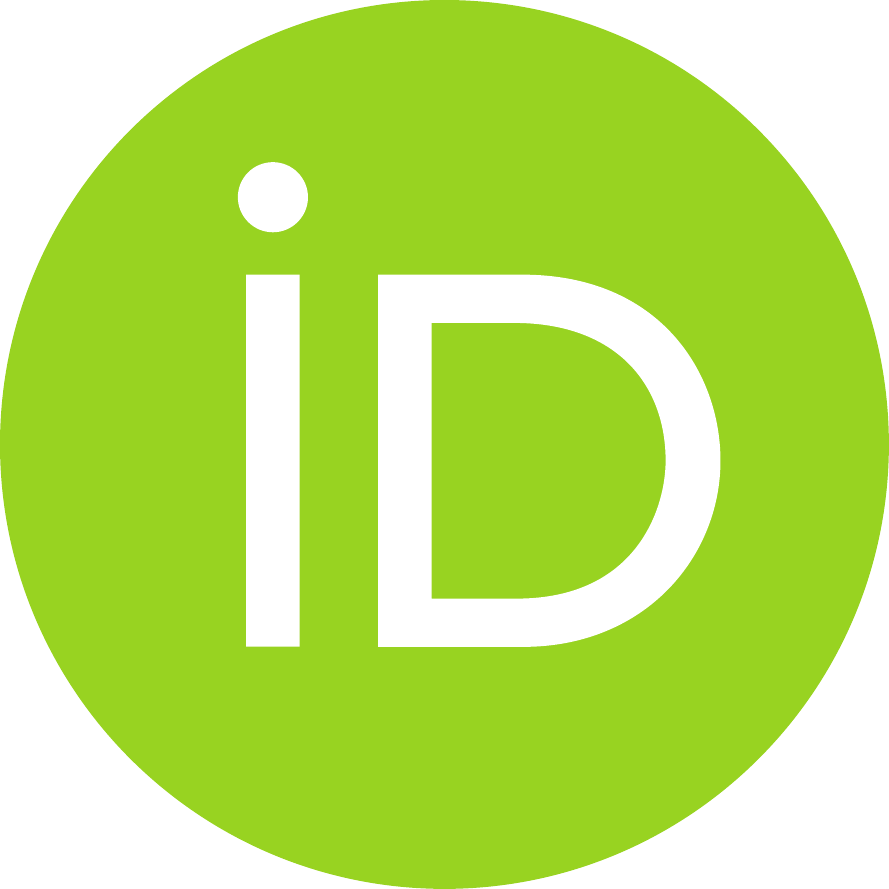}}}

\begin{document}


\title{Decoration of graphene nanoribbons by $5d$ transition-metal elements}

\author{Wei-Bang Li\,\orcidicon{0000-0002-8319-3316}}
\email[E-mail: ]{weibang1108@gmail.com}
\affiliation{Department of Physics, National Cheng Kung University, Tainan 70101, Taiwan}

\author{Kuang-I Lin\,\orcidicon{0000-0002-6487-1484}}
\email[E-mail: ]{kilin@mail.ncku.edu.tw}
\affiliation{Core Facility Center, National Cheng Kung University, Tainan 70101, Taiwan}

\author{Yu-Ming Wang\,\orcidicon{0000-0001-8866-065X}}
\email[E-mail: ]{wu0h96180@gmail.com}
\affiliation{Department of Physics, National Cheng Kung University, Tainan 70101, Taiwan}

\author{Hsien-Ching Chung\,\orcidicon{0000-0001-9364-8858}}
\email[E-mail: ]{hsienching.chung@gmail.com}
\affiliation{RD Dept., Super Double Power Technology Co., Ltd., Changhua City, Changhua County, 500042, Taiwan}

\author{Ming-Fa Lin}
\email[E-mail: ]{mflin@mail.ncku.edu.tw}
\affiliation{Department of Physics, National Cheng Kung University, Tainan 70101, Taiwan}

\date{\today}

\begin{abstract}
Graphene is a famous truly two-dimensional (2D) material, possessing a cone-like energy structure near the Fermi level and treated as a gapless semiconductor. Its unique properties trigger researchers to find applications of it. The gapless feature shrinks the development of graphene nanoelectronics. Making one-dimensional (1D) strips of graphene nanoribbons (GNRs) could be one of the promising routes to modulating the electronic and optical properties of graphene. The electronic and optical properties of GNRs are highly sensitive to the edge and width. The tunability in electronic and optical properties further implies the possibilities of GNR application. However, the dangling bonds at ribbon edges remain an open question in GNR systems. Various passivation at the ribbon edge might change the essential physical properties. In this work, $5d$ transition-metal elements are considered as the guest atoms at the edges. The geometric structure, energy bands, density of states, charge distribution, and optical transitions are discussed.
\begin{description}
\item[Usage]
This is a preprint version.
\end{description}
\end{abstract}

\maketitle


\section{Introduction}

A new scientific frontier has arrived since the discovery of graphene. This material triggers both fundamental scientists and applied technology engineers for many reasons. Graphene is the first truly two-dimensional material \cite{Science306(2004)666K.S.Novoselov, Proc.Natl.Acad.Sci.U.S.A.102(2005)10451K.S.Novoselov}, showing a cone-like energy spectrum near the Fermi energy, and is treated as a gapless semiconductor. The unique electronic structure of graphene result in many fascinating essential properties, such as high carrier mobility at room temperature ($>200,000$ cm$^2$/Vs) \cite{Science312(2006)1191C.Berger, SolidStateCommun.146(2008)351K.I.Bolotin, Phys.Rev.Lett.100(2008)016602S.V.Morozov}, superior thermoconductivity (3,000--5,000 W/mK) \cite{NanoLett.8(2008)902A.A.Balandin, Phys.Rev.Lett.100(2008)016602S.V.Morozov}, high transparency to incident light over a wide range of wavelength (97.7\%) \cite{Nat.Nanotechnol.5(2010)574S.Bae, Science320(2008)1308R.R.Nair}, as well as high modulus ($\sim$1 TPa) and tensile strength ($\sim$130 GPa) \cite{Science321(2008)385C.Lee}. Due to these superior electronic, thermal, mechanical, and optical properties, graphene is served as a high potential candidate for synthesizing next-generation electronic and optical devices.

The gapless feature results in a low on/off ratio in graphene-based field-effect transistors (FETs) and shrinks the development of graphene nanoelectronics. One of the most possible approaches to tune electronic and optical properties is making one-dimensional (1D) strips of graphene, i.e., graphene nanoribbons (GNRs) \cite{PhysicaE42(2010)711H.C.Chung, Phys.Chem.Chem.Phys.18(2016)7573H.C.Chung}. There are tremendous fabrication routes for GNR production, including both bottom-up and top-down schemes. From the geometric point of view, graphene cutting seems to be an intuitive route to fabricate GNRs, and the other available routes include lithographic patterning and etching of graphene \cite{NanoLett.9(2009)2083J.W.Bai, PhysicaE40(2007)228Z.H.Chen, Phys.Rev.Lett.98(2007)206805M.Y.Han, Nat.Nanotechnol.3(2008)397L.Tapaszto}, sonochemical breaking of graphene \cite{Science319(2008)1229X.L.Li, Phys.Rev.Lett.100(2008)206803X.R.Wang, NanoRes.3(2010)16Z.S.Wu}, oxidation cutting of graphene \cite{Adv.Mater.21(2009)4487L.Ci, Chem.Mater.19(2007)4396M.J.McAllister}, and metal-catalyzed cutting of graphene \cite{NanoLett.9(2009)2600L.C.Campos, Adv.Mater.21(2009)4487L.Ci, NanoRes.1(2008)116L.Ci, NanoLett.8(2008)1912S.S.Datta, Phys.StatusSolidiB246(2009)2540F.Schaffel, NanoRes.2(2009)695F.Schaffel, NanoLett.9(2009)457N.Severin, NanoRes.3(2010)16Z.S.Wu}. An interesting route is carbon nanotubes (CNTs) unzipping, since a CNT can be treated as a folded or zipped GNR \cite{BookLinRichQuasiparticlePropertiesLowDimensionalSystems}. The available routes for the reverse process contain chemical attack \cite{Carbon48(2010)2596F.Cataldo, Nature458(2009)872D.V.Kosynkin}, laser irradiation \cite{Nanoscale3(2011)2127P.Kumar}, plasma etching \cite{NanoRes.3(2010)387L.Jiao, Nature458(2009)877L.Jiao}, metal-catalyzed cutting \cite{NanoLett.10(2010)366A.LauraElias, Nanoscale3(2011)3876U.K.Parashar}, hydrogen treatment and annealing \cite{ACSNano5(2011)5132A.V.Talyzin}, intercalation and exfoliation \cite{NanoLett.9(2009)1527A.G.Cano-Marquez, ACSNano5(2011)968D.V.Kosynkin}, electrochemical unzipping \cite{J.Am.Chem.Soc.133(2011)4168D.B.Shinde}, sonochemical unzipping \cite{Nat.Nanotechnol.5(2010)321L.Jiao, J.Am.Chem.Soc.133(2011)10394L.Xie}, unzipping functionalized CNTs by scanning tunneling microscope (STM) tips \cite{NanoLett.10(2010)1764M.C.Pavia}, and electrical unwrapping by transmission electron microscopy (TEM) \cite{ACSNano4(2010)1362K.Kim}. Other routes include chemical vapor deposition (CVD) \cite{NanoLett.8(2008)2773J.Campos-Delgado, Nat.Nanotechnol.5(2010)727M.Sprinkle, J.Am.Chem.Soc.131(2009)11147D.C.Wei} and chemical synthesis \cite{ACSNano3(2012)2020S.Blankenburg, Nature466(2010)470J.M.Cai, J.Am.Chem.Soc.130(2008)4216X.Y.Yang, Appl.Phys.Lett.105(2014)023101Y.Zhang}. The former is piecewise linking of molecular precursor monomers, and the latter is much compatible with the current semiconductor industry.

The electronic and optical properties of GNRs are dominated by the ribbon width and the edge orientation. Zigzag GNRs (ZGNRs) exhibit partial flat subbands near the Fermi level with peculiar edge states localized at the ribbon edges \cite{J.Phys.Soc.Jpn.65(1996)1920M.Fujita, Phys.Rev.B54(1996)17954K.Nakada}, and the energy gaps of armchair GNRs (AGNRs) scale inversely with the ribbon width \cite{NanoLett.6(2006)2748V.Barone, Phys.Rev.Lett.97(2006)216803Y.W.Son}. The former are identified by the STM image \cite{Phys.Rev.B73(2006)125415Y.Kobayashi, Phys.Rev.B71(2005)193406Y.Kobayashi} and the latter have been confirmed by the electric conductance measurements \cite{Phys.Rev.Lett.98(2007)206805M.Y.Han, Science319(2008)1229X.L.Li} and tunneling current measurements \cite{Nat.Nanotechnol.3(2008)397L.Tapaszto}. On the other hand, the edge-dependent absorption selection rules of GNRs are predicted, i.e., $|\Delta n| = odd$ for ZGNRs and $\Delta n = 0$ for AGNRs, where $n$ is the subband index \cite{Opt.Express19(2011)23350H.C.Chung, Phys.Rev.B76(2007)045418H.Hsu, Phys.Rev.B84(2011)085458K.Sasaki}. These fundamental properties can be enriched by external fields, such as magnetic and electric fields.

A uniform static magnetic field perpendicular to the ribbon plane can accumulate the neighboring electronic states, inducing highly degenerate Landau levels (LLs) with quantized cyclotron orbits \cite{Z.Physik64(1930)629L.Landau}. The competition between the lateral confinement and the magnetic confinement enriches the magneto-electronic structures, e.g., partial dispersionless quasi-Landau levels (QLLs), 1D parabolic subbands, as well as partial flat subbands \cite{Phys.Rev.B73(2006)195408L.Brey, Phys.Rev.B59(1999)8271K.Wakabayashi}. Meanwhile, the magneto-optical spectra exhibit many symmetric and asymmetric absorption peaks. The symmetric absorption peaks result from the inter-QLL transitions and obey the magneto-optical selection rule of $|\Delta m| = 1$, where $m$ is an integer. However, the asymmetric absorption peaks originate from the transitions among parabolic subbands and abide by the edge-dependent selection rules \cite{Phys.Chem.Chem.Phys.18(2016)7573H.C.Chung}. A transverse static electric field generates extra potential energy in GNRs, i.e., the charge carriers experience different site energies \cite{J.Phys.Soc.Jpn.80(2011)044602H.C.Chung, Philos.Mag.94(2014)1859H.C.Chung}. The electronic and optical properties are drastically modified. The different potential energies in GNRs restrict the formation of Landau orbits, and QLLs would tilt, become oscillatory, or exhibit crossings and anti-crossings. Moreover, the inter-QLL optical transitions will be severely changed or even destroyed entirely \cite{Phys.Chem.Chem.Phys.18(2016)7573H.C.Chung, Carbon109(2016)883H.C.Chung, Phys.Chem.Chem.Phys.15(2013)868H.C.Chung}.

The edge-decorated GNRs, which cover the transition-metal guest adatoms, will be an interesting research strategy due to very diversified orbital hybridizations and spin configurations. The ten active orbitals of $5d$ can dominate the significant compounds, with special functionalities. But the concise quasi-particle mechanisms are almost absent in the previous studies. They will be achieved from the 1D transition-metal-decorated GNRs, including armchair and zigzag systems. The $d$-orbitals are enriched by the spin-up and spin-down configurations. Their probability distributions are characterized by the $6s$ and $5d$ orbitals, which include $d_{xy}$, $d_{yz}$, $d_{z^2}$, $d_{xz}$, and $d_{x^2}$, being enriched by the spin-up and spin-down configurations. The spatial electron density, which mainly comes from the linear superposition of $6s$ and $5d$ orbitals, will be very sensitive to the chemical environments. In this work, the guest adatoms, Tantalum ($_{73}$Ta with atomic number $Z=73$ and electron configuration [Xe]$4f^{14}5d^36s^2$) and Tungsten ($_{74}$W with atomic number $Z=74$ and electron configuration [Xe]$4f^{14}5d^46s^2$), possess strong reactions with dangling carbon bonds at two open edges of ZGNRs and AGNRs. The pristine systems are predicted to display, respectively, the ferromagnetic and anti-ferromagnetic spin configurations on the same sides and across the ribbon center and the fully non-magnetic properties. How to modulate and diversify magnetic behaviors through transition-metal decorations should be a very interesting research topic.

\section{Geometric structures for transition metal-decorated graphene nanoribbons}

\begin{figure}[]
  \includegraphics[keepaspectratio]{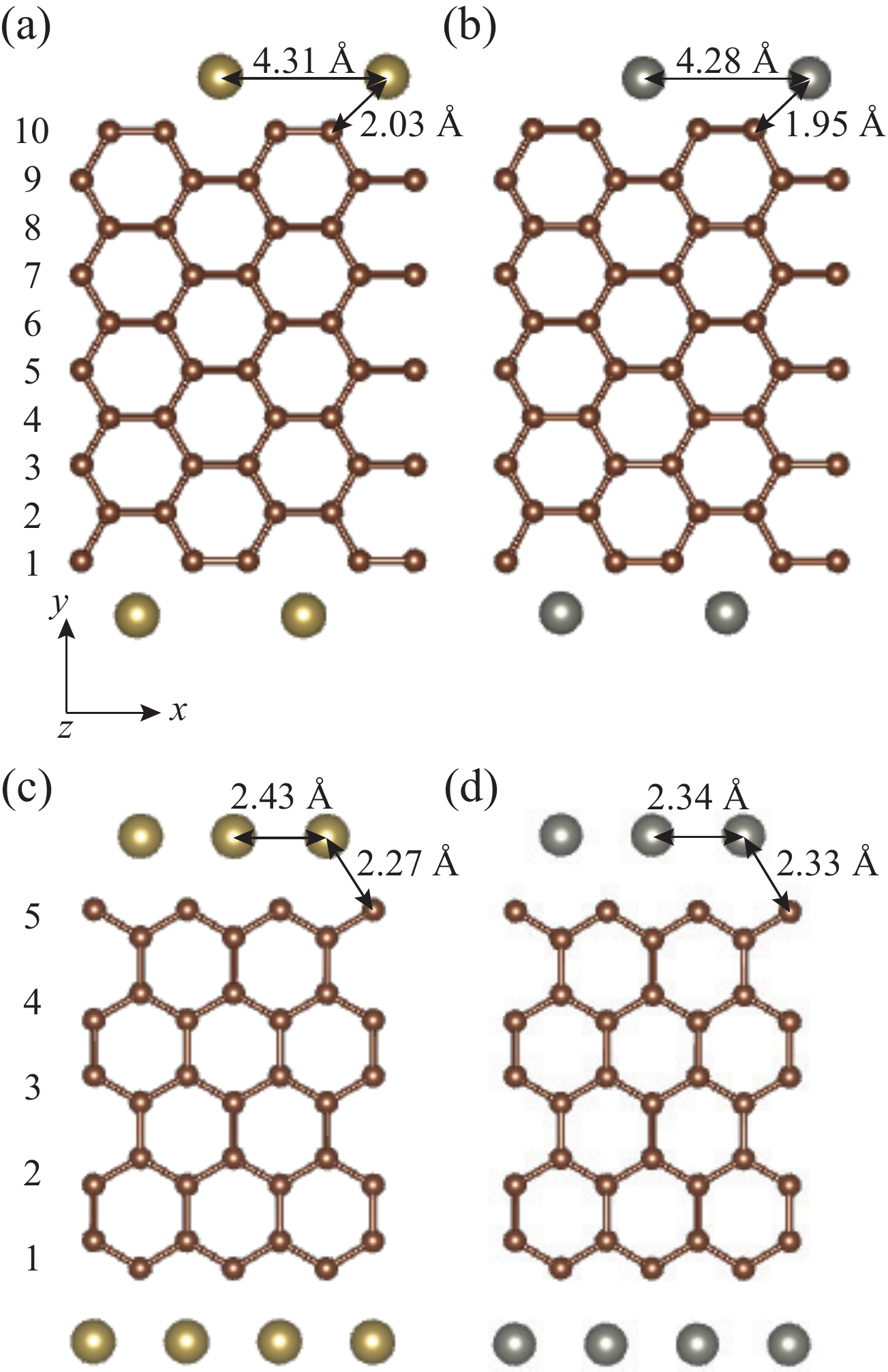}\\
  \caption{
(Color online)
The geometry structure of (a) Ta- and (b) W-decorated AGNRs with $N_A = 10$, and (c) Ta- and (d) W-decorated ZGNRs with $N_Z = 5$. The brown, yellow, and gray balls represent the carbon (C), Tantalum (Ta), and Tungsten (W) atoms, respectively. The numbers label the $i$-th dimer and zigzag lines.
}
  \label{fig:Figure01}
\end{figure}

The atomic numbers of $_{73}$Ta and $_{74}$W, which belong to the transition metal elements, are chosen for the unusual quasi-particles. All Vienna \emph{ab}-initio Simulation Package (VASP) calculations are finished under delicate analyses to achieve concise pictures of physics, chemistry, and material science. The VASP simulations are very suitable for fully exploring the unusual quantum quasi-particles of Ta- and W-decorated 1D AGNRs and ZGNRs. After the significant decorations, their critical roles in all essential properties are thoroughly examined and identified from the consistent results. We established the framework under the unified correlation among the multi-orbital hybridizations, spin configurations, crystal structures, electronic properties, and optical properties. We can find there exist no buckling structures on the Ta- and W-decorated lattices via C--Ta and C--W bonds, as shown in Fig.~\ref{fig:Figure01}.

The AGNRs and ZGNRs are good candidates for edge-decoration studies because of the unsaturated bonds. The widths of AGNRs ($N_A$) and ZGNRs ($N_Z$) are, respectively, determined by the number of dimer lines and zigzag lines along the periodic direction ($x$-direction). In our case, the $N_A$ and $N_Z$ are chosen as 10 and 5, respectively. The C--C bond length is about 1.42 {\AA} (similar to the value in graphene), but it possesses fluctuations within 1.42--1.48 {\AA} in GNRs. The further the carbon atoms are from the edge, the less fluctuations the bond lengths will be~\cite{arXiv:2206.11162}. Such feature results in the highly non-uniform physical environment and thus the sensitive changes of $2p_z$--$2p_z$ hopping integrals. The edge structures are totally different before and after the chemical Ta- and W-decorations. The non-uniform edge structures appear on both AGNRs and ZGNRs. For the former, the Ta--Ta and W--W bond lengths are 4.31 {\AA} and 4.28 {\AA}, respectively. The C--Ta and C--W bond lengths are 2.03 {\AA} and 1.95 {\AA}, respectively. Moreover, the C--C bond lengths are reduced to a small fluctuation within 1.40--1.43 {\AA}. These reflect that the structures become more stable due to the edge decorations. The C--W bond length seems stronger than the C--Ta ones. For the zigzag cases, the Ta--Ta and W--W bond lengths are 2.43 {\AA} and 2.34 {\AA}, respectively. The C--Ta and C--W bond lengths are 2.27 {\AA} and 2.33 {\AA}, respectively. The C--Ta bond lengths are shorter than the C--W ones, this indicates that the C--Ta are more stable in zigzag systems probably due to the interactions within W atoms.

\section{Energy band structures and density of states}

\begin{figure*}[]
  \includegraphics[keepaspectratio,height=21cm]{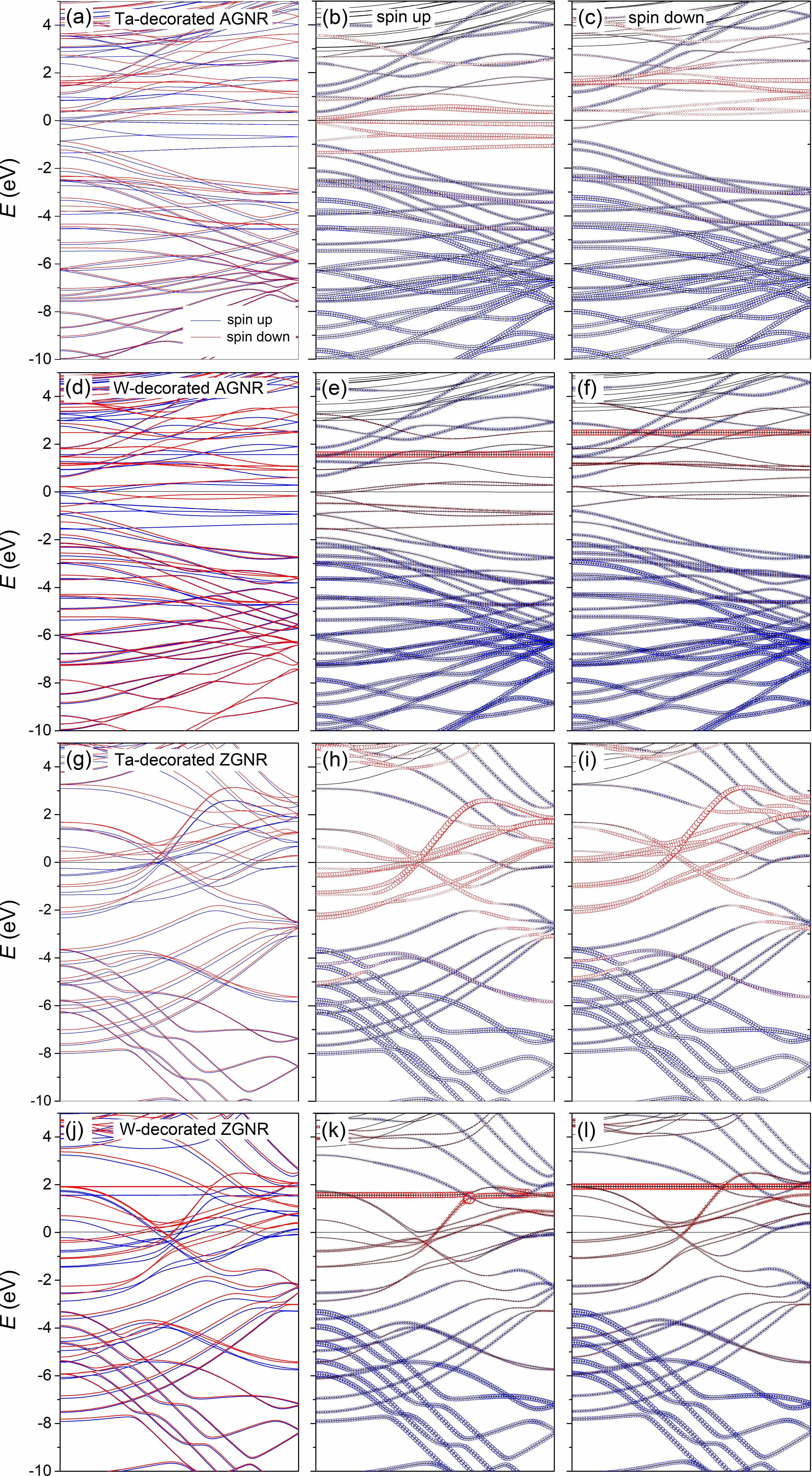}\\
  \caption{
(Color online)
(a) The spin-splitting energy band structures of Ta-decorated AGNR, where the spin-up and spin-down are indicated by blue and red curves, respectively.
(b) The spin-up band structure of Ta-decorated AGNR, where the red and blue circles indicate Ta-atom- and C-atom-dominated states, respectively.
(c) The spin-down band structure of Ta-decorated AGNR.
(d) The spin-splitting energy band structures of W-decorated AGNR.
(e) The spin-up band structure of W-decorated AGNR, where the red and blue circles indicate W-atom- and C-atom-dominated states, respectively.
(f) The spin-down band structure of W-decorated AGNR.
(g) (h) (i) For Ta-decorated ZGNR.
(j) (k) (l) For W-decorated ZGNR.
}
  \label{fig:Figure02}
\end{figure*}

There are a lot of differences between graphene, few-layered graphene, graphene-related intercalation compounds, and GNR systems. It was known that there exists a pair of partial flat valance band and conduction band. Such electronic wave functions are localized near the open zigzag edges. Therefore, they belong to the edge localization states with almost vanishing group velocities. Band gaps decline as ribbon widths grow. However, the past researches exhibit that the armchair systems possess three various  energy bands with respect to various groups of $N_A$, where $N_A=3I$, $3I+1$, and $3I+2$ and $I$ is a positive integer. In our work, the low-lying energy bands, as clearly illustrated in Fig.~\ref{fig:Figure02}, indicate the dramatic changes after the edge decorations of transition metal atoms. There are much more subbands, and they are dominated by C atoms, W atoms or Ta atoms, or co-dominated by C--Ta or C--W bondings. The asymmetric valence and conduction energy subbands are greatly enhanced through the close partnerships of guest-host and host-host interactions. The apparent changes of the band-edge states cover their numbers, energies, critical points, and curvatures. Moreover, the energy band structures can exhibit spin-split behaviors, mainly owing to the $5d$ orbitals of the transition metal guest atoms, where their splitting energies should be comparable to the on-site spin-dependent Coulomb interactions. Most of the bands, which are not cross-section with the Fermi energy ($E_F$), are fully occupied or fully unoccupied.
So the net contributions of magnetic moments of them are almost zero. However, there are few bands across the $E_F$, and they will dominate the magnetic properties. The magnetic momenta of spin-up and spin-down states will display the distinct contributions to the net magnetic moment. That is to say, the edge-decorated GNRs possess net magnetic momenta not equal to zero when the spin-split energy bands are under different occupied states, e.g., the spin-up energy band is fully occupied but the spin-down one is just partial occupied. In both the AGNRs and ZGNRs, the orbitals of carbon atoms-($2s$, $2p_x$, $2p_y$, $2p_z$) are slightly changed but the spin-split bands of Ta atoms-($6s$, $5d_{xy}$, $5d_{yz}$, $5d_{z^2}$, $5d_{xz}$, $5d_{x^2}$) are apparently separated from each other within -2.0--2.0 eV in the armchair system. So does the W atom case. The spin-split orbitals of carbon atoms remain the similar states to each other in armchair and zigzag systems, but the spin-split bands near the $E_F$ of W atoms are very different between spin-up and spin-down cases. The first-principles simulations are available for predicting the edge-decorated GNRs. For establishing a complete quasi-particle framework, the calculated predictions need to be examined by the experiments, such as angle-resolved photoemission spectroscopy (ARPES), which can determine the subband-dependent and spin-dependent Fermi momenta.

\begin{figure}[t]
  \includegraphics[keepaspectratio]{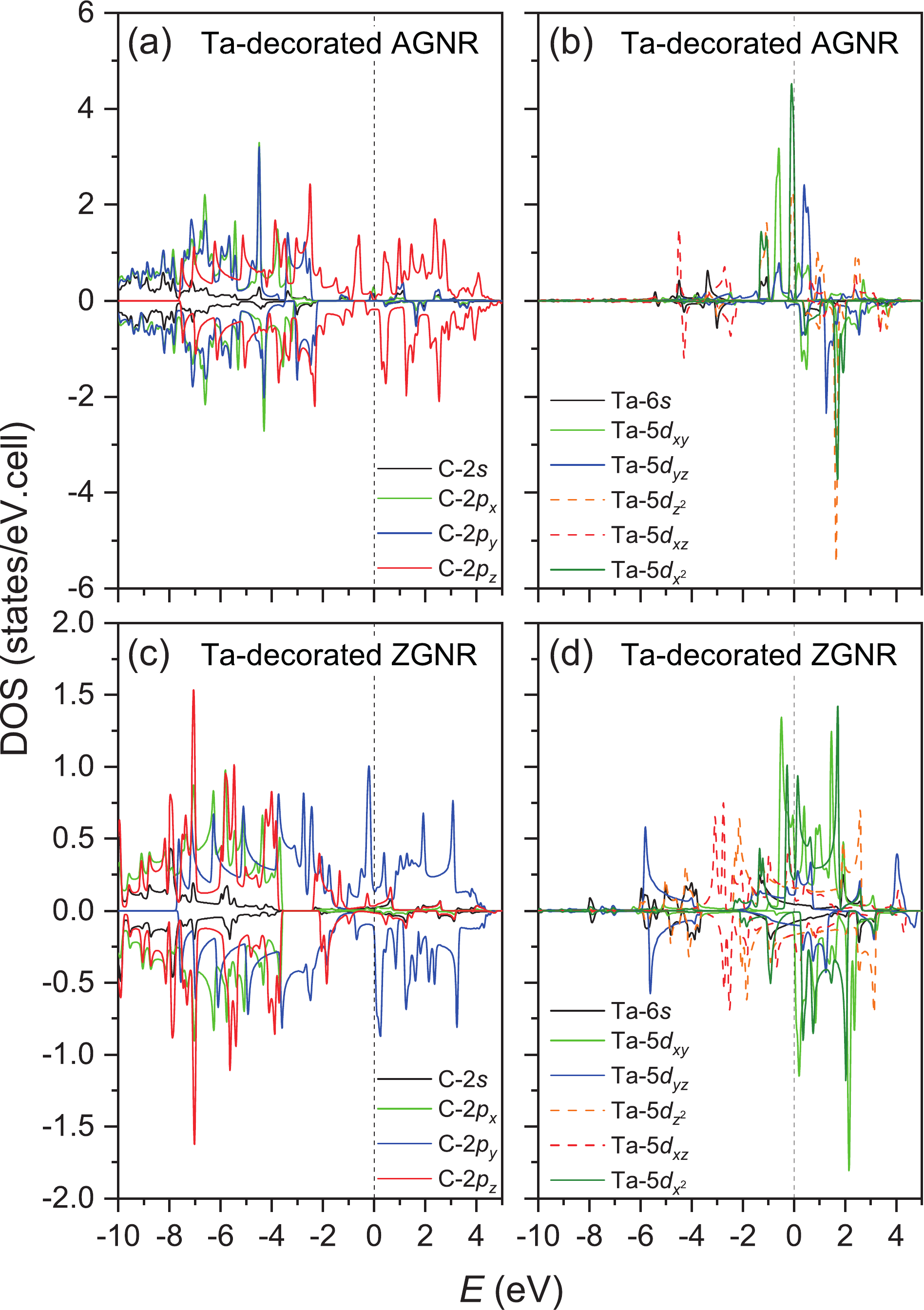}\\
  \caption{
(Color online)
The partial density of states of (a) C and (b) Ta atoms in AGNRs and the partial density of states of (c) C and (d) Ta atoms in ZGNRs. The dashed line indicates $E_F=0$.
}
  \label{fig:Figure03}
\end{figure}

\begin{figure}[t]
  \includegraphics[keepaspectratio]{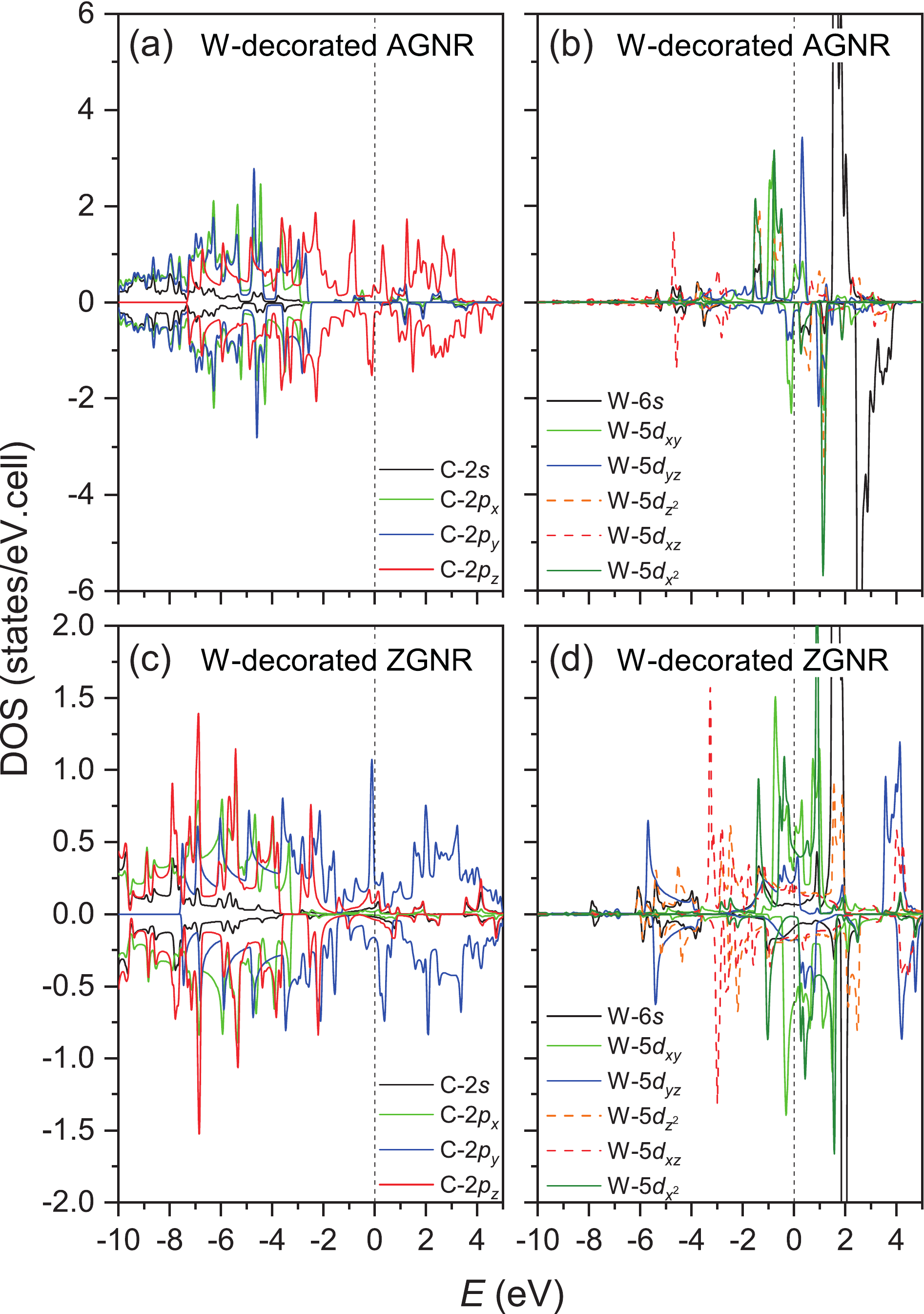}\\
  \caption{
(Color online)
The partial density of states of (a) C and (b) W atoms in AGNRs and the partial density of states of (c) C and (d) W atoms in ZGNRs. The dashed line indicates $E_F=0$.
}
  \label{fig:Figure04}
\end{figure}

There exist highly symmetric points (critical points) in the energy wave-vector space  and they can represent the van Hove singularities. For one dimension, e.g., the GNR systems, the density of states (DOS) can be obtained from the constant energy surface. Fig.~\ref{fig:Figure03} displays the energy-dependent DOSs through the atom- and orbital-decomposed spin-split contributions, which include the C atoms and Ta atoms in the armchair system (Figs.~\ref{fig:Figure03}(a) and (b)), and in the zigzag system (Figs.~\ref{fig:Figure03}(c) and (d)); and includes the C atoms and W atoms in the armchair system (Figs.~\ref{fig:Figure04}(a) and (b)), and in the zigzag system (Fig.~\ref{fig:Figure04}(c) and (d)). In the armchair systems, the C-($2s$, $2p_x$, $2p_y$) exist at $E < 2.5$ eV, $-0.8$ eV $< E < -0.4$ eV, and $0.3$ eV $< E < 0.4$ eV. There exist spin-split structures mainly on C-$2p_z$ within $-1$ eV $< E < 1$ eV; in the zigzag systems, the C-($2_s$, $2p_x$, $2p_y$) appears at $E < -3.5$ eV, and $-2.4$ eV $< E < 4.2$ eV, and the split of C-$2p_z$ are apparent within $-1$ eV $< E < 1$ eV. However, in the armchair systems, the spin-split orbitals of Ta and W atoms exist at a wide range of $-1.8$ eV $< E < 4$ eV; in the zigzag systems, the spin-split orbitals of Ta and W atoms exist at $-1.5$ eV $< E < 5$ eV. According to the van Hove singularities, C-$2p_z$ and Ta-$5d$ co-dominate weakly at $-5$ eV $< E < -4$ eV, and $-3$ eV $< E < -2$ eV, and co-dominate strongly at -0.8 eV, 0.5 eV, and 0.6 eV for armchair case, and these reveal the existence of C--Ta bonding. Similar to the Ta-decorated case, the C-$2p_z$ and W-$5d$ strongly co-dominate at -0.9 eV, -0.1 eV, 0.3 eV, and -1 eV.

For the zigzag system, there are many van Hove singularities near the Fermi level, and the C-$2p_z$ and Ta-$5d$ co-dominate at $-1$ eV $< E < 1$ eV, and 2 eV; and the Ta-$5d$ dominates at -0.8 eV, and 1.4 eV, and the Ta-$6s$ dominate at 0.2 eV and 1.7 eV. These reveal that the C--C, C--W or C--Ta, and W--W or Ta--Ta must co-exist in the edge-decorated GNRs. However, the van Hove singularities of W-$6s$ orbitals appear much strongly at $1$ eV $< E < 2$ eV. This might be caused by unsaturated W-$6s$ orbitals, in which one of the electrons in $6s$ will be transferred to the $5d$ orbitals to achieve the half-filled constructions of $5d$. Also, according to the symmetry of lattice, the guest atoms are much closer to each other than in the armchair ones. Thus, the W-atoms will interact with not only the C atoms, but also the nearest W atoms to form the metallic bonding. In the W-decorations of zigzag systems, the flat bands at 1.6 eV and 1.9 eV are dominated by the W-$6s$ and in good agreement with the van Hove singularities density of states; in the armchair case, the flat bands exist at 1.6 eV and 2.6 eV and accord with the van Hove singularities as well. According to the energy band structures and density of states, we find that there co-exist the multi-orbital hybridizations of C--C, C--W, C--Ta, W--W, and Ta--Ta, which correspond to
($2s$, $2p_x$, $2p_y$, $2p_z$)--($2s$, $2p_x$, $2p_y$, $2p_z$),
($2s$, $2p_x$, $2p_y$, $2p_z$)--($6s$, $5d_{xy}$, $5d_{yz}$, $5d_{z^2}$, $5d_{xz}$, $5d_{x^2}$),
($2s$, $2p_x$, $2p_y$, $2p_z$)--($6s$, $5d_{xy}$, $5d_{yz}$, $5d_{z^2}$, $5d_{xz}$, $5d_{x^2}$),
($6s$, $5d_{xy}$, $5d_{yz}$, $5d_{z^2}$, $5d_{xz}$, $5d_{x^2}$)--($6s$, $5d_{xy}$, $5d_{yz}$, $5d_{z^2}$, $5d_{xz}$, $5d_{x^2}$), and
($6s$, $5d_{xy}$, $5d_{yz}$, $5d_{z^2}$, $5d_{xz}$, $5d_{x^2}$)--($6s$, $5d_{xy}$, $5d_{yz}$, $5d_{z^2}$, $5d_{xz}$, $5d_{x^2}$), respectively. The orbital hybridization of C--C also covers the $2p_z$--$2p_z$ near the Fermi level.

\section{Charge distributions, charge variations, and optical properties}

\begin{figure*}[h]
  \includegraphics[keepaspectratio]{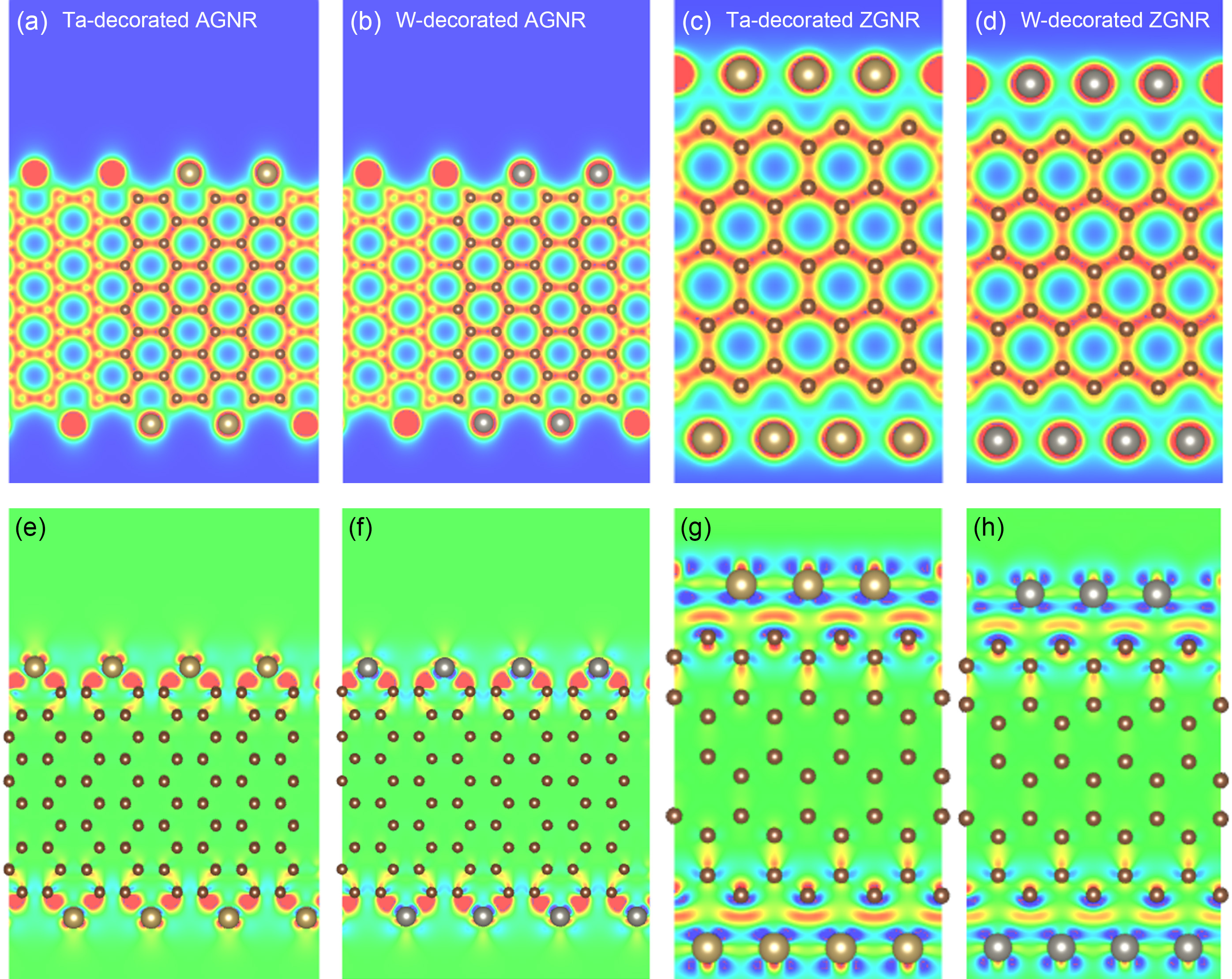}\\
  \caption{
(Color online)
The spatial charge distribution of (a) Ta-decorated AGNR, (b) W-decorated AGNR, (c) Ta-decorated ZGNR, and (d) W-decorated ZGNR, respectively.
The charge variations of (e) Ta-decorated AGNR, (f) W-decorated AGNR, (g) Ta-decorated ZGNR, and (h) W-decorated ZGNR, respectively.
}
  \label{fig:Figure05}
\end{figure*}
The edge-decorations cause the non-uniform chemical environment in the Ta- and W-decorated GNRs. The spatial charge distributions and variations of one-dimension edge-decorated systems are shown in Fig.~\ref{fig:Figure05}. The carbon atoms, Ta atoms, and W atoms exhibit the distribution of spherical probability. The red region and the green-yellow region in the spatial charge distributions, respectively, represent the $2s$ and ($2p_x$, $2p_y$, $2p_z$) orbitals for carbon atoms; the red region and the green-yellow region exhibit $6s$ and ($5d_{xy}$, $5d_{yz}$, $5d_{z^2}$, $5d_{xz}$, $5d_{x^2}$) orbitals for Ta or W atoms as well.
The calculated results display that the C--C bonding corresponding to the red region is very strong, and there exists weak but significant bonding on the edged atoms, which are covered by the green region. It reveals that the W and Ta atoms form the orbital hybridization mainly with the $2p_z$ of the near carbon atoms. Very interesting, the charge variations show the differences between the armchair and zigzag systems. In the zigzag system, the lightly yellow regions exist between the guest atoms after decoration, but seldom in the armchair systems. That is to say, the metallic bonding is more apparent in the zigzag edge in the zigzag systems. However, the red region between the carbon atom and guest atom is wider in the armchair systems than in the zigzag systems, and this reveals that the carbon-metal bonding is stronger in the armchair systems. These special distributions are caused due to the geometric structures, in other words, the guest atoms are far from each other in the AGNRs, but close in the zigzag ones. These complicated mixing of orbital hybridizations are in agreement with the energy band structures and van Hove singularities of the DOSs.

\begin{figure}[]
  \includegraphics[]{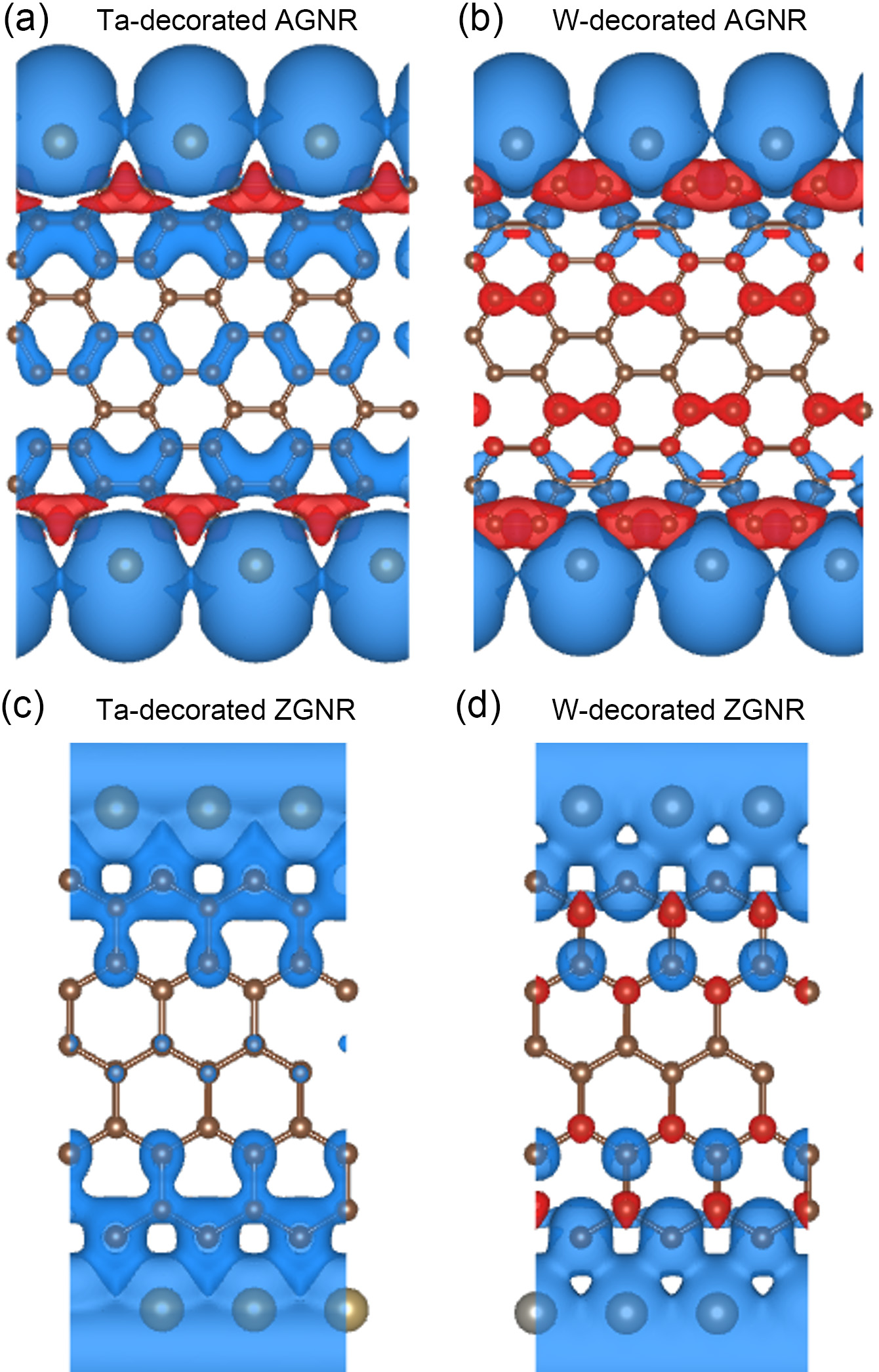}\\
  \caption{
(Color online)
The spatial spin distributions of (a) Ta- and (b) W-decorated AGNR and (c) Ta- and (b) W-decorated ZGNR, respectively. The spin-up states and spin-down states are indicated by blue and red colors, respectively.
}
  \label{fig:Figure06}
\end{figure}

We use the orbital-projected net magnetic moments to analyze the distributions of the density of spin, as shown in Fig.~\ref{fig:Figure06}. The transition-metal atoms can make great contributions ferromagnetic spin configurations at the edges due to the $5d$ orbitals. The spin-up states and spin-down states represent, respectively, the blue region and red region in Fig.~\ref{fig:Figure06}. The results display much weak spin-down (red region) in the Ta-, W-decorated AGNRs and W-decorated ZGNRs. However, there exist almost spin-up states in the Ta-decorated zigzag systems, which correspond to the van Hove singularities in the spin-split DOSs. Combining with the calculated energy band structures and density of states, the spin interaction might appear between the guest atoms, in other words, the interactions of ($5d_{xy}$, $5d_{yz}$, $5d_{z^2}$, $5d_{xz}$, $5d_{x^2}$)--($5d_{xy}$, $5d_{yz}$, $5d_{z^2}$, $5d_{xz}$, $5d_{x^2}$) exist even though the bond lengths of them are longer than guest-carbon bond lengths.

\begin{figure}[]
  \includegraphics[]{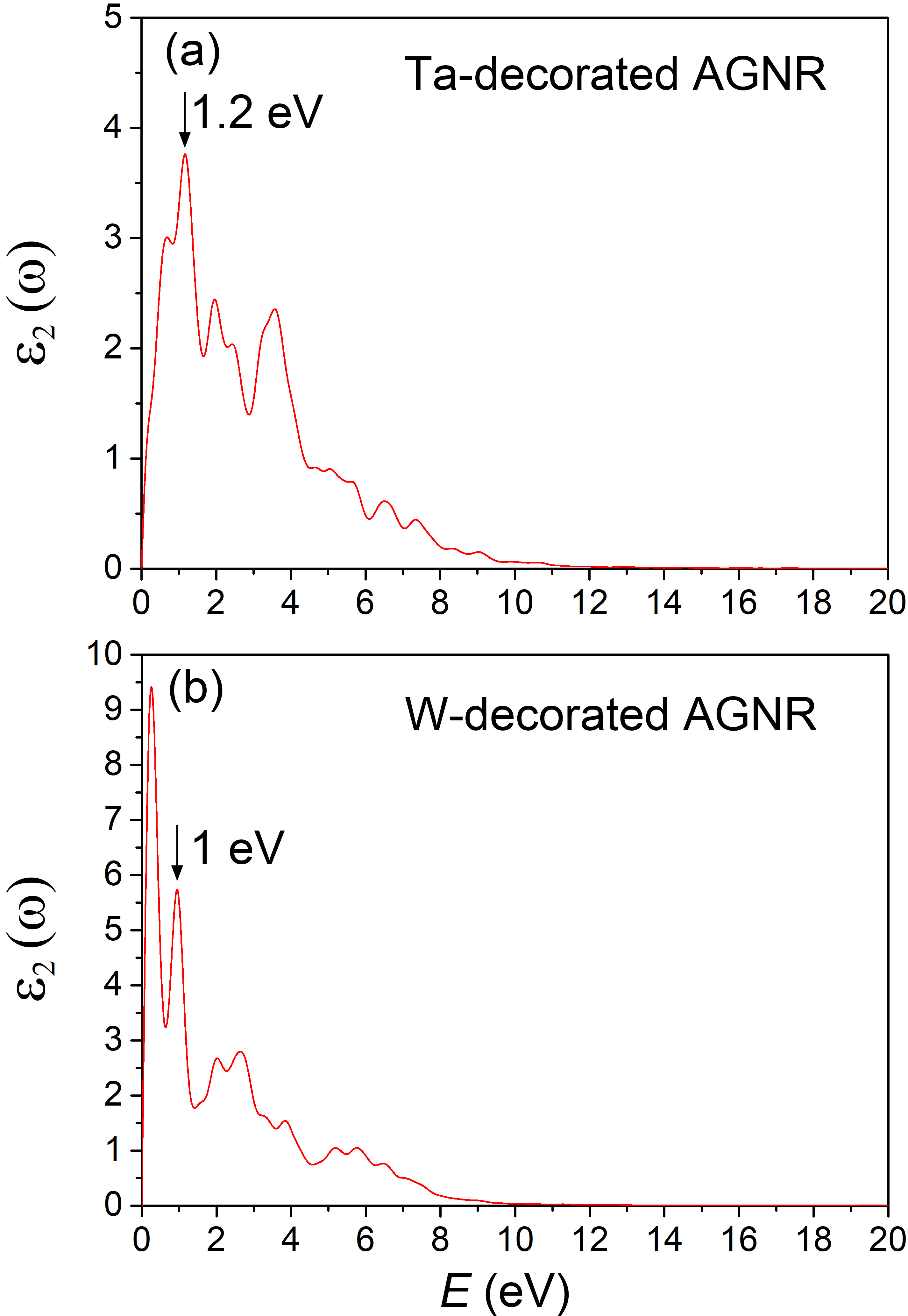}\\
  \caption{
(Color online)
The imaginary parts of the dielectric functions of (a) Ta- and (b) W-decorated AGNRs, respectively.
}
  \label{fig:Figure07}
\end{figure}

Our work shows the optical properties with the imaginary part of the dielectric function, which represent the photon-electron couplings under the requirements of momentum and energy conservations and the spin-dependent Pauli exclusion principle. In the armchair systems of Ta- and W-decorations, there presents a threshold peak in the imaginary-prat dielectric function which arises from the part of $5d$-orbitals about the complicated initial and final states. The pure $\pi$-electronic excitations, which possess comparable spin-up and spin-down contributions, come to exist as the second absorption peak at 1.2 eV and 1 eV for Ta- and W-decorated systems, respectively. As shown in Fig.~\ref{fig:Figure07}. The other prominent absorption peaks belong to the composite quasi-particle pictures, in which they consist of the superposition about C-$2p_z$, C-($2s$, $2p_x$, $2p_y$, $2p_z$), C-($2s$, $2p_x$, $2p_y$), and part of Ta- and W-orbitals. These unusual atom-, orbital-, and spin-dominated optical quasi-particles clearly illustrate the diversified excitation phenomena, being very successful in developing an enlarged framework. The composite quasi-particles are well characterized by their intrinsic properties, further illustrating an enlarged framework in combination with other chapters. The high-precision optical measurements of absorption and transmission spectroscopies are available in observing the rich single-particle and collective excitations. It is very interesting and important to investigate these properties through experiments in the future.

\section{Conclusion}

The first-principles calculations with VASP are very useful and successful in investigating the quasi-particles via the geometric structures, charge distributions and variations, spin-split energy band structures and density of states, and spatial spin distributions. The transition-metal atoms, which possess $5d$ orbitals, make great contributions to the compounds in low-lying energy, such as a lot of many merged structures of the atom-, orbital- and spin-projected van Hove singularities, the atom- and orbital-decomposed magnetic moments, the Ta-/W-induced spin arrangements, and the spin-enriched optical transitions in dielectric functions. The critical properties in the transition-metal-edged decorations mainly arise from the orbital hybridizations of host-guest atoms, host-host atoms, and guest-guest atoms. That is to say, the bonding of C--Ta, C--W, Ta--Ta, W--W, and C--C arise from
($2s$, $2p_x$, $2p_y$, $2p_z$)--($6s$, $5d_{xy}$, $5d_{yz}$, $5d_{z^2}$, $5d_{xz}$, $5d_{x^2}$),
($2s$, $2p_x$, $2p_y$, $2p_z$)--($6s$, $5d_{xy}$, $5d_{yz}$, $5d_{z^2}$, $5d_{xz}$, $5d_{x^2}$),
($6s$, $5d_{xy}$, $5d_{yz}$, $5d_{z^2}$, $5d_{xz}$, $5d_{x^2}$)--($6s$, $5d_{xy}$, $5d_{yz}$, $5d_{z^2}$, $5d_{xz}$, $5d_{x^2}$),
($6s$, $5d_{xy}$, $5d_{yz}$, $5d_{z^2}$, $5d_{xz}$, $5d_{x^2}$)--($6s$, $5d_{xy}$, $5d_{yz}$, $5d_{z^2}$, $5d_{xz}$, $5d_{x^2}$), and
($2s$, $2p_x$, $2p_y$, $2p_z$)--($2s$, $2p_x$, $2p_y$, $2p_z$), respectively. In addition, the bonding of C--C covers $2p_z$--$2p_z$. Moreover, according to the spatial spin distributions, the interaction of spin just appears between transition-metal atoms no matter in armchair or zigzag systems. Based on the calculated results, the C-$2p_z$ and five $5d$ orbitals almost co-dominate the physical and chemical properties in transition-metal-decorated GNRs within the low-lying energy. In summary, the quasi-particle properties, including spin-split orbitals and optical properties, are seldom done in the past years. It is very difficult to fully and accurately investigate the transition-metal-based compounds due to the five complicated $5d$ orbitals-($5d_{xy}$, $5d_{yz}$, $5d_{z^2}$, $5d_{xz}$, $5d_{x^2}$). For further exploration and construction of the full framework of quasi-particles, it is worthwhile to combine the first-principles simulations with the phenomenological models, and experimental methods, such as high-resolution ARPES and spin-polarized scanning tunneling spectroscopy (STS), in the future.

The commercialization of electric vehicles (EVs) worldwide \cite{engrxiv2020OutlooksLiIonBatteriesH.C.Chung, BookChChung2021EngIntePotentialAppOutlooksLiIonBatteryIndustry, GlobalEVOutlook2020IEA} results in the mass production of lithium-ion (Li-ion) batteries. The retired power batteries of EVs have largely increased, causing resource waste threats. Recycling and utilization of such retired batteries have been suggested \cite{Sustain.EnergyTechnol.Assess.6(2014)64L.Ahmadi, Renew.Sustain.EnergyRev.93(2018)701E.Martinez-Laserna, CellRep.Phys.Sci.2(2021)100537J.Zhu}, since many retired power batteries still possess about 80\% initial capacity \cite{J.Environ.Manage.232(2019)354L.C.Casals, FMEAofLFPBatteryModule2018Chung, WorldElectr.Veh.J.9(2018)24A.Podias, J.EnergyStorage11(2017)200S.Tong, J.PowerSources196(2011)5147E.Wood}. They can be repurposed once again, serving as the battery modules in the energy storage system \cite{Batteries3(2017)10L.C.Casals, BookChChung2021UL1974, EnergyPolicy71(2014)22C.Heymans, WasteManage.113(2020)497D.Kamath, Batteries5(2019)33H.Quinard}. Governments have noticed this serious problem and are prepared to launch policies to deal with the recovery and reuse of repurposing batteries \cite{Sci.Data8(2021)165H.C.Chung, J.TaiwanEnergy6(2019)425H.C.Chung, MJTE860(2020)35H.C.Chung, BookChChung2021UL1974, EnergyPolicy113(2018)535K.Gur, IEEEAccess7(2019)73215E.Hossain}. The discovery of graphene and more low-dimensional materials (such as 1D finite-width GNRs discussed in this work) \cite{Sci.Rep.9(2019)2332H.C.Chung, BookLinFirstPrinciplesCathodeElectrolyteAnodeBatteryMaterials} makes scientists believe that these materials can be potential additives in battery materials to improve battery performance and to overcome the recycling problems of Li-ion batteries.

\section*{Acknowledgements}

The author (H. C. Chung) thanks Pei-Ju Chien for English discussions and corrections as well as Ming-Hui Chung, Su-Ming Chen, Lien-Kuei Chien, and Mi-Lee Kao for financial support. This work was supported in part by Super Double Power Technology Co., Ltd., Taiwan, under the project “Development of Cloud-native Energy Management Systems for Medium-scale Energy Storage Systems (\href{https://osf.io/7fr9z/}{https://osf.io/7fr9z/})” (Grant number: SDP-RD-PROJ-001-2020).
This work was supported in part by Ministry of Science and Technology (MOST), Taiwan (grant numbers: MOST 111-2112-M-006-020, MOST 109-2124-M-006-001, and MOST 110-2634-F-006-017).
This work was supported in part by National Science and Technology Council (NSTC), Taiwan (grant number: NSTC 111-2811-M-006-047).

\bibliography{Reference}
\bibliographystyle{apsrev4-2}

\end{document}